\documentclass[twocolumn]{aastex63}
\received{June 30, 2020}
\revised{July 31, 2020}
\accepted{August 10, 2020}
\submitjournal{AJ}

\shorttitle{Near-infrared characterization of four massive stars in transition phases}
\shortauthors{Cochetti et al.}

\begin{document}

\title{Near-infrared characterization of four massive stars in transition phases\footnote{Based on observations obtained 
1) at the Gemini Observatory, which is operated by the Association of Universities for Research in Astronomy, Inc., under a cooperative agreement with the NSF on behalf of the Gemini partnership: the National Science Foundation (United States), the National Research Council (Canada), CONICYT (Chile), the Australian Research Council (Australia), Minist\'erio da Ci\^encia, Tecnologia e Inova\c c\~ao (Brazil) and Ministerio de Ciencia, Tecnolog\'ia e Innovaci\'on Productiva (Argentina), under programs: GN-2013A-Q-78 and GN-2014A-Q-36, 
2) with the Nordic Optical Telescope, operated by the Nordic Optical Telescope Scientific Association at the Observatorio del Roque de los Muchachos, La Palma, Spain, of the Instituto de Astrofisica de Canarias, under program IDs 59-417 and 61-102 (OPTICON proposal 20A/011).
3) with ESO Telescopes at the La Silla Paranal Observatory under program IDs 384.D-1078(A) and 088.D-0442(B), and 
4) at Tartu Observatory of Tartu University, Estonia}
}

\correspondingauthor{Yanina R. Cochetti}
\email{cochetti@fcaglp.unlp.edu.ar}

\author[0000-0002-2763-7250]{Yanina R. Cochetti}
\affiliation{Departamento de Espectroscop\'ia, Facultad de Ciencias Astron\'omicas y Geof\'isicas, Universidad Nacional de La Plata\\
Paseo del Bosque S/N, La Plata, B1900FWA, Buenos Aires, Argentina}
\affiliation{Instituto de Astrof\'isica de La Plata (CCT La Plata - CONICET, UNLP)\\
Paseo del Bosque S/N, La Plata, B1900FWA, Buenos Aires, Argentina}

\author[0000-0002-4502-6330]{Michaela Kraus}
\affiliation{Astronomical Institute, Czech Academy of Sciences, Fri\v{c}ova 298, 251\,65 Ond\v{r}ejov, Czech Republic\\}

\author{Mar\'ia L. Arias}
\affiliation{Departamento de Espectroscop\'ia, Facultad de Ciencias Astron\'omicas y Geof\'isicas, Universidad Nacional de La Plata\\
Paseo del Bosque S/N, La Plata, B1900FWA, Buenos Aires, Argentina}
\affiliation{Instituto de Astrof\'isica de La Plata (CCT La Plata - CONICET, UNLP)\\
Paseo del Bosque S/N, La Plata, B1900FWA, Buenos Aires, Argentina}

\author{Lydia S. Cidale}
\affiliation{Departamento de Espectroscop\'ia, Facultad de Ciencias Astron\'omicas y Geof\'isicas, Universidad Nacional de La Plata\\
Paseo del Bosque S/N, La Plata, B1900FWA, Buenos Aires, Argentina}
\affiliation{Instituto de Astrof\'isica de La Plata (CCT La Plata - CONICET, UNLP)\\
Paseo del Bosque S/N, La Plata, B1900FWA, Buenos Aires,  Argentina}

\author{T\~onis Eenm\"ae}
\affiliation{Tartu Observatory, University of Tartu, 61602 T{\~o}ravere, Tartumaa, Estonia\\}

\author[0000-0003-2196-9091]{Tiina Liimets}
\affiliation{Astronomical Institute, Czech Academy of Sciences, Fri\v{c}ova 298, 251\,65 Ond\v{r}ejov, Czech Republic\\}
\affiliation{Tartu Observatory, University of Tartu, 61602 T{\~o}ravere, Tartumaa, Estonia\\}

\author{Andrea F. Torres}
\affiliation{Departamento de Espectroscop\'ia, Facultad de Ciencias Astron\'omicas y Geof\'isicas, Universidad Nacional de La Plata\\
Paseo del Bosque S/N, La Plata, B1900FWA, Buenos Aires, Argentina}
\affiliation{Instituto de Astrof\'isica de La Plata (CCT La Plata - CONICET, UNLP)\\
Paseo del Bosque S/N, La Plata, B1900FWA, Buenos Aires,  Argentina}

\author[0000-0001-6316-9880]{Anlaug A. Djupvik}
\affiliation{Nordic Optical Telescope, Rambla Jos\'e Ana Fern\'andez P\'erez 7, ES-38711 Bre\~na Baja, Spain}



\begin{abstract}

Massive stars typically undergo short-lived post-main sequence evolutionary phases with strong mass loss and occasional mass eruptions. Many of such massive stars in transition phases have been identified based on their dusty envelopes. The ejected material often veils the stellar photospheres so that the central stars cannot be assigned proper spectral types and evolutionary stages. The infrared spectral range has proved to be ideal for the classification of evolved massive stars and for the characterization of their environments. To improve our knowledge on the central stars of four such dust enshrouded objects: [GKF2010]~MN~83,  [GKF2010]~MN~108, [GKF2010]~MN~109, and  [GKF2010]~MN~112, we collect and present their first medium resolution K-band spectra in the $2.3\,-\,2.47\,\mu$m region and discuss the location of the stars in the JHK color-color diagram. We find that the emission-line spectra of both MN~83 and MN~112 show characteristics typically seen in Luminous Blue Variable (LBV) stars. In addition, we propose that the presence and strength of the newly reported Mg~{\sc ii} lines might be used as a new complementary criterion to identify LBV candidates. The spectra of the other two objects imply that MN~108 is an O-type supergiant, whereas MN~109 could be an LBV candidate in its active phase. We derive lower limits for the reddening toward the stars and find that three of all de-reddened fall into the region of confirmed LBVs.

\end{abstract}



\section{Introduction} \label{sec:intro}

For most of their life massive stars interact with their environments through high-density stellar winds. They add chemically processed material into the interstellar medium (ISM) and modify the dynamics of their neighborhood depositing momentum and energy. After massive stars leave the main sequence, they can evolve through individual short-lived transition phases of intense mass loss or eruption events before ending their lives as supernovae. Moreover, as the stellar evolution is sensitive to the mass loss \citep{Smith2014}, the amount of mass lost in each stage affects the subsequent evolution of the star \citep{Maeder1980, Meynet2005, Georgy2012}. Despite their importance on the fate of massive stars, some short transition phases are not reproducible with the currently available evolution models. In this context, to improve our comprehension of stellar and galactic evolution, a better understanding of massive stars in transition phases is vital \citep{Langer1994, Puls2008, Langer2012}. 

As a result of the high mass loss or mass eruptions, massive stars in transition phases usually present large amounts of circumstellar material in the form of disks, shells or rings, with physical conditions that allow for the formation of dust and molecules. This circumstellar material often hides the stellar photosphere, so that it can be difficult to assign proper spectral types and to determine the evolutionary stage of the central star. Nevertheless, the material around the star provides information about the kinematics and the history of the mass loss through specific atomic and molecular emission features.

One class of transition objects is comprised of the luminous blue variables \citep[LBVs, see][for a definition of the phenomena]{Conti1997}. Their progenitors are very massive and luminous stars. However, the physical state of these stars prior to or post the LBV phase is not well constrained, but seems to depend on the initial mass of the star. While the most massive stars ($M > 40$\,M$_{\odot}$) seem to develop into LBVs shortly after they left the main sequence, their lower mass counterparts ($M < 40$\,M$_{\odot}$) seem to turn into LBVs only after they have passed through the red-supergiant phase \citep{2011BSRSL..80..266M}.

LBVs possess relatively slow dense winds, with high mass loss rates. Moreover, most LBVs are surrounded by a nebula, presenting a wide range of morphologies, from ring-like shells to bipolar lobes and triple-ring systems \citep{Nota1995, Weis2001, Smith2007}. These nebulae are most likely formed from material ejected either 
during a giant eruption, or from phases of enhanced mass loss during the so-called S\,Dor cycles. Starting from a regular blue supergiant (BSG) stage (quiescent state), the star apparently develops into a cool A or F-type supergiant (outburst). During such an S\,Dor outburst, which can last from years to decades 
\citep{1997A&AS..124..517V}, the mass loss rate can be increased by up to a factor of five \citep{2001A&A...375...54S, 2008RMxAC..33..132G} before the star finally returns into its quiescence state as regular blue supergiant. Only if a star has experienced a giant eruption or a full S\,Dor cycle it is classified as LBV, otherwise only as a candidate LBV (cLBV). 

Another group in transition comprises the B[e] supergiants (B[e]SGs). These luminous ($4<\log\,L/L_{\sun}<6$) emission-line stars form a subset of the early-type supergiants. Stars within this group are known for their peculiar spectral character suggesting a complex circumstellar environment. Their optical spectra display intense and broad emission lines of the Balmer series along with narrow emission lines of neutral and singly ionized metals in both permitted and forbidden transitions \citep{Cannon1924, Henize1956, Smith1957, Feast1960}. Moreover, these objects possess an intense near-infrared excess due to hot circumstellar dust \citep{Allen1976, Glass1977, Stahl1983, Stahl1984}. The observed peculiarities can be understood in terms of a coexistence of high-ionized polar winds, also traced by the broad blueshifted UV resonance lines of highly ionized elements, and equatorial rings or disks \citep{Zickgraf1985}. The circumstellar rings/disks consist of low-ionized or neutral atomic material \citep{Kraus2007,Kraus2010,Aret2012}, hot molecular gas, and warm dust \citep{McGregor1988,Morris1996,Kastner2006,Kastner2010,Meilland2010, Wheelwright2012,Kraus2013}. This disk material is in quasi-Keplerian rotation and clearly detached from the central star \citep{Aret2012,Cidale2012,deWit2014,Aret2016,Kraus2017,Krausetal2017,Maravelias2018}. 

In contrast to the LBVs, most B[e]SGs are not reported to possess large-scale ejecta. The only exceptions so far are the three Galactic objects MWC\,137 \citep{Marston2008, Krausetal2017}, MWC\,314 \citep{Marston2008}, and MWC\,349A \citep{2012A&A...541A...7G, 2020MNRAS.tmp..491K}.

These two different groups of stars (LBVs and B[e]SGs) occupy similar regions in the HR diagram \citep{Aret2012}, and to assign a star a classification as either LBV or B[e]SG is not trivial. Candidates for these two groups are identified through spectroscopic features or circumstellar structures similar to the ones associated with these types of evolved massive stars \citep[e.g.][]{Clark2003, Clark2012}. However, optical and IR features observed in the spectra of B[e]SGs and LBVs in quiescence are practically indistinguishable \citep{Oksala2013}, and only subtle details could be used for a proper classification. For instance, emission from [O~{\sc i}] and [Ca~{\sc ii}] lines, or CO bands are characteristic for B[e]SGs \citep[][see also Section \ref{sec:dis}]{Kraus2019}. In addition, these two classes of stars can be separated based on their distinct locations in infrared color-color diagrams \citep{Oksala2013, Kraus2019}. However, the lack of photometric measurements along with full spectral coverage often hampers a proper stellar classification. This is especially true for the large number of recently identified evolved massive stars embedded within dense infrared shells \citep{Wachter2010, GKF2010, G2010, G2012}.

In this work we study four objects with infrared shells, listed in \citet{GKF2010} and proposed to host either a BSG or a cLBV as central source. To re-investigate and to improve their classification, we present and analyze the medium resolution $K$-band spectra of these objects. As has been shown by many authors \citep[see, e.g.,][]{2014ApJ...780L..10K, 2014MNRAS.443..947L, Arias2018, Kourniotis2018}, spectral features in the near-infrared provide valuable complementary information for the classification of massive stars in transition phases. In addition, we present optical spectra and photometric measurements for two of them. Optical narrow-band images are obtained for all four, and near-infrared images for two targets. Our goal is to obtain supplementary information about the circumstellar material that can help to define the possible evolutionary states of the central objects.

\section{The selected sample}
\label{sec:lit}

From the catalog of \citet{GKF2010} of evolved massive stars with infrared nebulae, we selected the four northern objects which have been proposed to host either a BSG or a cLBV as central source. In this section, we briefly describe these objects based on the literature status.

{\bf [GKF2010]~MN~83:} Using images from Spitzer IRAC and VLA, \citet{Davies2007} found a bright ring nebula in $8~\mu$m and 20~cm, respectively, with a highly reddened star at its center (see their Figure~21). These authors associated the nebula to the source $IRAS\,18367-0556$ \citep{Becker1994}, and remarked that this object is very similar to the cLBV HD~168625 \citep{Robberto1998}. The bulk of the radio emission at 1.4 GHz reported by \citet{Condon1998} is in the southwestern part of the dust ring, rather than aligned with the central star. Then, it is improbable that the radio emission comes from the ionized stellar wind. Instead, the radio emission might be originated from the ionization of the surrounding ejecta because of the presence of a hot central star, or from a fast wind pushing a slower, dusty wind ejected in a cooler phase of the star.

\citet{GKF2010} reported an ellipsoidal nebula with a size of $30\arcsec\times 45\arcsec$ in the MIPS 24-$\mu$m images, with a radio counterpart at 20 cm. These authors included this source in their catalog as MN~83 and classified it as a cLBV, based on the work of \citet{Davies2007}.

Using Chandra observations, \citet{Naze2012} confirmed the non-detection of X-ray emission from MN~83. The closest X-ray source is the BSG GAL~026.47+00.02 situated at a distance of $\sim30\arcsec$.  

{\bf [GKF2010]~MN~108:} \citet{GKF2010} discovered a bipolar nebula on Spitzer MIPS images with a size of $50\arcsec\times 50\arcsec$. These authors also reported strong optical absorption lines of H~{\sc i} and He~{\sc i}, and numerous absorption lines of He~{\sc ii}, being the strongest  He~{\sc ii}~$\lambda$4686 and He~{\sc ii}~$\lambda$5412~\AA. They classified this object as a BSG star.

{\bf [GKF2010]~MN~109:} \citet{PhillipsRamos2008} identified a ring-like structure around the star (which they named G052.4-00.0 in their paper) based on images from the Spitzer GLIMPSE mid-IR survey of the galactic plane and the MIPS 24~$\mu$m images. This ring-like structure has not been detected either in the IRAS survey or in DSS images. Neither have radio measurements been performed so far. The central star is seen, however, in the near-IR Two-Micron All-Sky Survey (2MASS), based on which \citet{PhillipsRamos2008} concluded that the central star is an early-type supergiant at a distance of $\sim8$\,kpc from the Sun. It is associated with nearby regions of star formation. The radius of the infrared ring is of the order of $\approx0.3$\,pc, and the extinction $A_{\rm V}$ toward the central star is of the order of $21$\,mag. \citet{GKF2010} included  MN~109 in their catalog as a BSG candidate with an ellipsoidal nebula having a size of $15\arcsec\times 20\arcsec$. 

{\bf [GKF2010]~MN~112:} This object is listed in the catalog of H$\alpha$ emission stars in the northern Milky Way \citep{Kohoutek1999}. Associated to this object, \citet{GKF2010} identified a round nebula with a diameter of $100\arcsec$ which is also visible at 24~$\mu$m \citep{Wachter2010}. Its morphology resembles that of the circumstellar nebula of the Galactic cLBV GAL 079.29+11.46. The optical spectrum of MN~112 is dominated by strong emission lines of H~{\sc i} and He~{\sc i}, and numerous lines of N~{\sc ii}, Fe~{\sc iii} and Si~{\sc ii}. Based on the observed H~{\sc i} and He~{\sc i} emission lines, \citet{Wachter2010} classified it as a B-type star ($\sim$B0-B5). Due to the resemblance of its optical spectrum to the one of the LBV star P\,Cyg \citep{G2010}, \citet{Stringfellow2012} and \citet{G2012} assign this object a cLBV status.

\section{Observations} \label{sec:obs}

\begin{deluxetable*}{l|cc|cr@{$\times$}lc|ccc}
\tablecaption{Observed stellar sample. Identification number, equatorial coordinates (right ascension $\alpha$ and declination $\delta$), and observing log. \label{table:obs}}
\tablecolumns{10}
\tablewidth{0pt}
\tablehead{
\multicolumn{1}{c|}{ID} &
\multicolumn{1}{c}{$\alpha$} &
\multicolumn{1}{c|}{$\delta$} &
\multicolumn{4}{c|}{IR Spectra} &
\multicolumn{3}{c}{Optical Spectra} \\
\multicolumn{1}{c|}{[GKF2010]} &
\multicolumn{1}{c}{[J2000]} &
\multicolumn{1}{c|}{[J2000]} &
\multicolumn{1}{c}{Obs. date} & \multicolumn{2}{c}{Exp. time~(sec)} & \multicolumn{1}{c|}{SNR} &
\multicolumn{1}{c}{Obs. date} & \multicolumn{1}{c}{Exp. time~(sec)} & \multicolumn{1}{c}{SNR} 
}
\startdata
MN~83  & 18:39:23.01 & $-$05:53:19.9 & 2014-05-09 &  8&200& 190 &...        &...  &...\\
MN~108 & 19:26:58.96 &   +18:46:44.0 & 2014-07-05 & 20&280& 120 & 2019-08-13 & 1800 & 40 \\
MN~109 & 19:28:14.58 &   +17:16:23.1 & 2014-03-20 &  8&630& 220 &...        &...  &...\\
MN~112 & 19:44:37.60 &   +24:19:05.9 & 2013-07-07 &  8&192& 270 & 2019-08-13 & 3600 & 100 \\
\enddata
\end{deluxetable*}

\subsection{Spectroscopy}

Infrared (IR) spectra in the K-band were obtained with the Gemini Near Infrared Spectrograph (GEMINI/GNIRS) in long-slit mode. The selected instrumental configuration was: a 111 l/mm grating, a 0.3 arcsec slit, and the short camera ($0\farcs15$\,pix$^{-1}$). This configuration provides a spectral resolution of R\,$\sim$\,6000 in the spectral range 2.29-2.48~$\mu$m. For each target, several ABBA sequences were taken. A late-B or early-A type star was observed near in time and sky position to perform the telluric absorption correction. Flats and arcs were taken with every science target. The data reduction steps include the subtraction of the AB pairs, flat-fielding, telluric correction and wavelength calibration. The reduction process was carried out with the {\sc IRAF}\footnote{IRAF is distributed by the National Optical Astronomy Observatory, which is operated by the Association of Universities for Research in Astronomy (AURA) under cooperative agreement with the National Science Foundation.} software package.

Optical spectroscopic observations were done with the 1.5-meter telescope AZT-12 at Tartu Observatory of Tartu University, Estonia, using the long-slit spectrograph ASP-32 with a 300 l/mm grating. This instrumental setting yields a spectral resolution of R\,$\sim$\,1000 in the 4000-7300~\AA\ spectral range. 

The spectroscopic observing log is given in Table~\ref{table:obs}.

\subsection{Photometry}

Photometric observations were carried out with the 0.31-meter Planewave CDK12.5 telescope at Tartu Observatory. The telescope is equipped with an Apogee Alta U42 CCD-camera and Astrodon Photometrics $B$, $V$, $R_{\rm C}$ and $I_{\rm C}$ Bessell filters. All observations were calibrated by zero, dark, and flat frames. For measurements and transformations to the Johnson-Cousins standard system, the AVSO's VPHOT service was used. Comparison stars were selected from the APASS\footnote{https://www.aavso.org/apass} survey data release 10. $R_{\rm C}$ and $I_{\rm C}$ magnitudes of the comparison stars were calculated from SLOAN $r$ and $i$ filter measurements of APASS, using relationships described in \citet{Jester2005}.

The exposure times for the photometric measurements in the four filters were 180, 120, and 60~sec for the stars MN~83, MN~108, and MN~112, respectively. MN~109 was observed purely in the $I_{\rm C}$ filter with an exposure time of 300~sec.

\subsection{Imaging}

Our objects were additionally observed with The Nordic Optical Telescope (NOT) using the Andalucia Faint Object Spectrograph and Camera (ALFOSC) between 2019 September 5 and 2020 June 20. The observations have been carried out with the narrow-band H$\alpha$ filter with a central wavelength 6577~\AA\ and a full width at half maximum of 180~\AA, covering the emission lines H$\alpha$~$\lambda$6562.82~\AA\, and [N~{\sc ii}]~$\lambda\lambda6548,~6583$~\AA. The pixel scale of ALFOSC is $0\farcs21$\,pix$^{-1}$, and the field of view (FOV) is $6\farcm4\times6\farcm4$. The ALFOSC data was processed (biased, flat-fielded, combined) using the standard routines in IRAF. The total exposure time for each target was 30 minutes.

In addition, NOT's near-IR Camera and spectrograph (NOTCam\footnote{See http://www.not.iac.es/instruments/notcam/ for info on NOTCam and for info related to data reduction packages in IRAF.\label{footnotelabel}}) was used with the Br$\gamma$ narrow-band filter (NOT \#209) centered at $2.163~\mu$m to map MN~83 on 2019 September 13  and MN~112 on 2020 June 24. The NOTCam detector is a Hawaii 1k HgCdTe IR array with which the wide-field camera gives a pixel scale of $0\farcs234$\,pix$^{-1}$ and a FOV of $4\arcmin\times4\arcmin$. The observations were obtained in beam-switch mode with an additional small-step dither due to the search for faint extended emission and the rather dense stellar fields. Every exposure consists of a pixel by pixel linear regression of a number of non-destructive reads, sampling the signal up the ramp until the exposure time is reached. For MN~83 we used 50s exposures and for MN~112 32s. Each exposure was flat-fielded, sky-subtracted, distortion corrected, and shifted using the NOTCam IRAF package\textsuperscript{\ref{footnotelabel}}. The sky was evaluated  with the OFF-source frames, and the flat-field was obtained from differential twilight flats. The final images were obtained by ``median" combining all exposures leading to total exposure times of 450 seconds for MN~83 and 576 seconds for MN~112.

To supplement our images, Spitzer's Infrared Array Camera (IRAC; \citealt{2004ApJS..154...10F}) images in all four bands (3.6, 4.5, 5.8, and 8.0~$\mu$m) and The Multiband Imaging Photometer for Spitzer (MIPS; \citealt{2004ApJS..154...25R}) in the two bands (24 and 70~$\mu$m) were downloaded from the NASA/IPAC Infrared Science Archive (IRSA). The IRAC images have a pixel scale $0\farcs6$\,pix$^{-1}$ and MIPS images $2\farcs45$\,pix$^{-1}$.

\section{Results} \label{sec:res}

\subsection{Spectroscopy}

\begin{figure*}[t!]
\includegraphics[width=\textwidth]{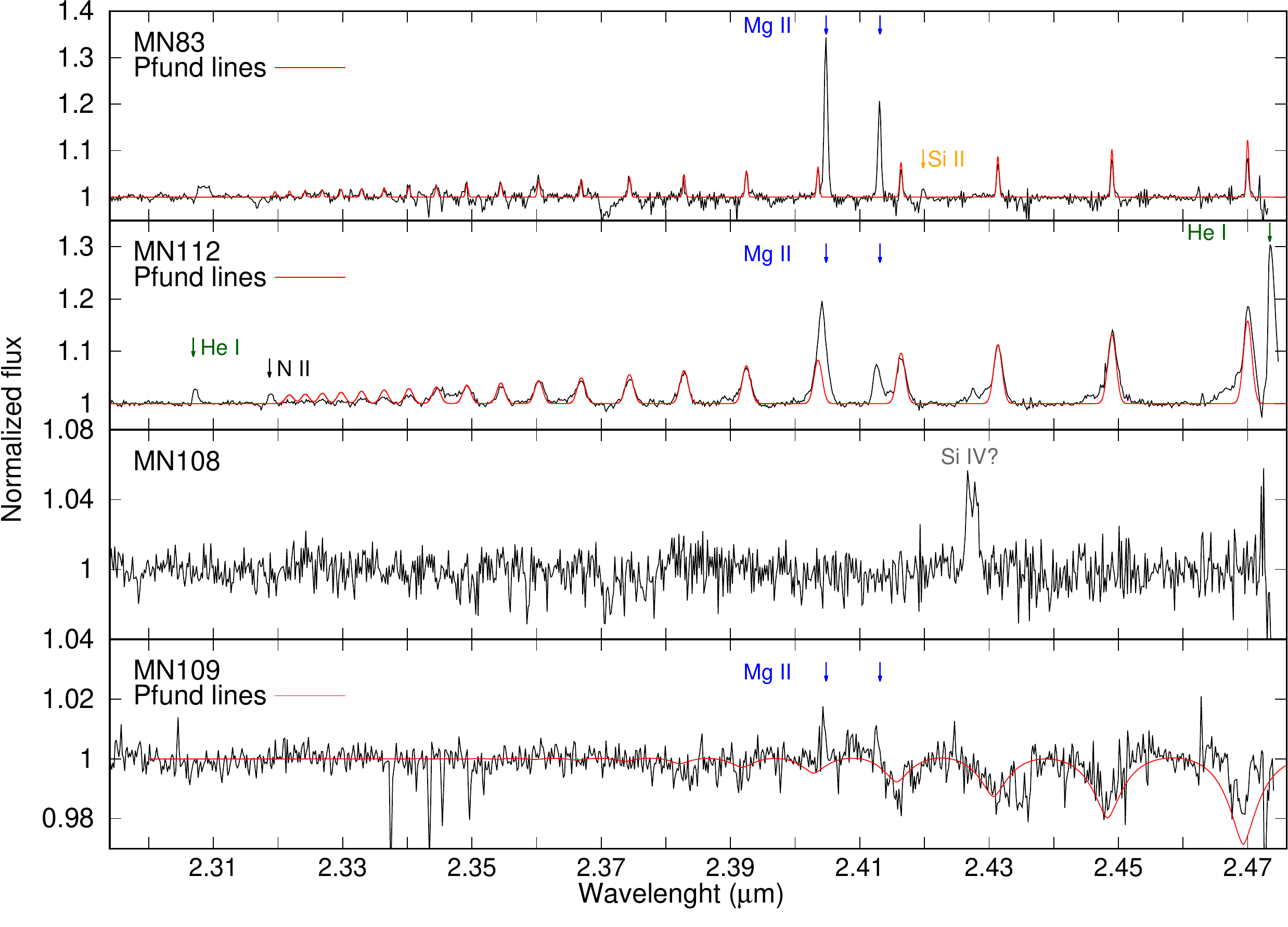}
\caption{$K$-band spectra from our stellar sample. The hydrogen Pfund series is in emission in MN~83 and MN~112 and in absorption in MN~109. Synthetic spectra (modeled with the code from \citet{Kraus2000} for MN~83 and MN~112; and from a \citet{Kurucz1979} model of a star with $T_{\rm eff}=10\,000$~K and $\log g=3.0$ for MN~109) are plotted in red. The arrows point out other intense spectral lines observed in this spectral range. 
\label{fig:Kband}}
\end{figure*}

\begin{figure*}[t!]
\includegraphics[width=\textwidth]{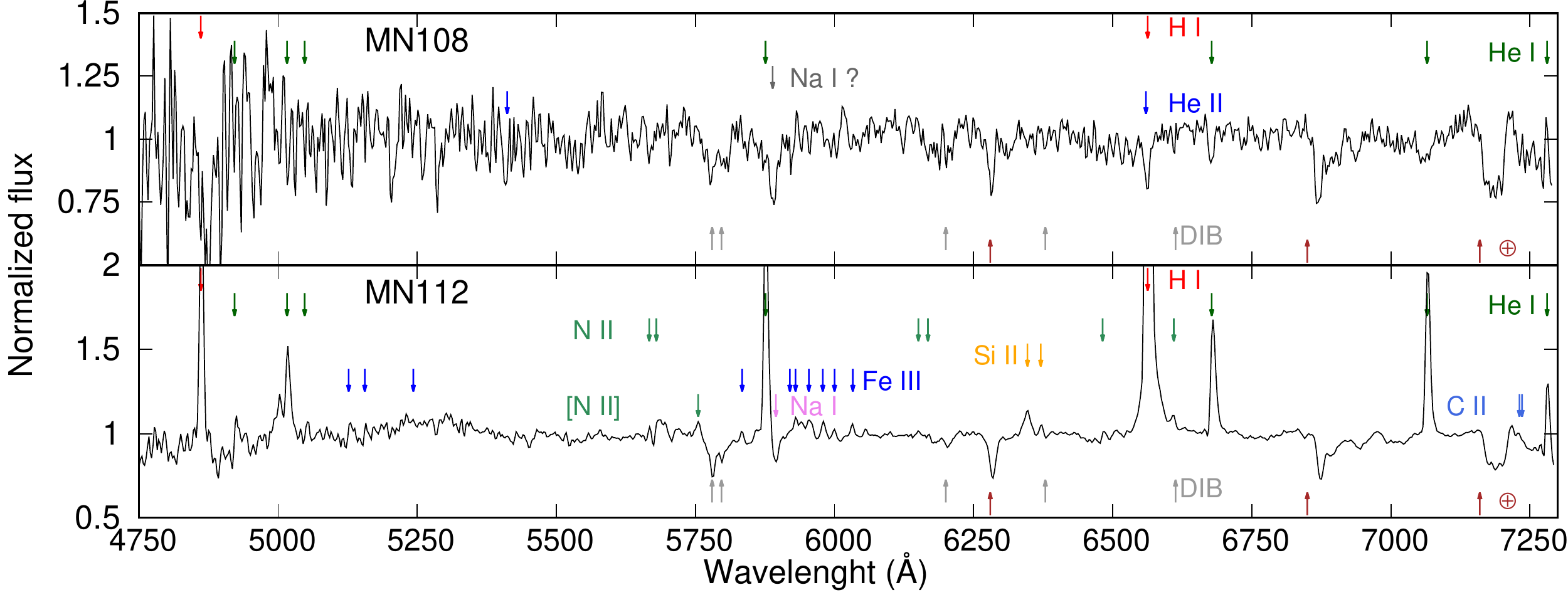}
\caption{Optical spectra of MN~108 and MN~112. The arrows point out the position of the most intense identified lines, together with the position of the identified diffuse interstellar bands (DIBs) and the telluric bands (pointed out with the symbol $\Earth$).} 
 \label{fig:optical}
\end{figure*}

We present our IR and optical spectra in Figures~\ref{fig:Kband} and \ref{fig:optical}, respectively, and the photometric measurements in Table~\ref{table:photometry}. For all objects, these are the first medium resolution K-band spectra in the $2.3\,-\,2.47\,\mu$m region that have been reported. 

Inspection of Figure~\ref{fig:Kband} reveals that, obviously, none of the stars displays CO bands, neither in emission nor in absorption.

The K-band spectra of MN~83 and MN~112 are similar at first glance. Both present emission from the hydrogen Pfund series and from the lines of Mg~{\sc ii} $\lambda\lambda2.4047,~2.4131~\mu$m. The latter are even more intense than the hydrogen lines. Despite these similarities, there are also differences between these two spectra. First, the emission lines in MN~83 are extremely narrow, suggesting that the emission in MN~83 might arise in a nebula. Second, the K-band spectrum of MN~112 also displays the lines of He~{\sc i} $\lambda2.3070~\mu$m and  He~{\sc i} $\lambda2.4734~\mu$m, and  an emission feature at $\lambda2.3187~\mu$m that could be N~{\sc ii}. The emission of He~{\sc i} lines suggests that the central star in MN~112 is most likely of spectral type O, whereas MN~83 seems to host a B-type central source. The latter object has only one extra feature, an emission line at $2.42~\mu$m, which might be identified as due to either Si~{\sc ii} $\lambda2.4199~\mu$m or Ar~{\sc ii} $\lambda2.4203~\mu$m. All other features are likely artifacts resulting from the extraction process and telluric remnants.

The hydrogen Pfund emission lines are associated with the presence of a hot, dense, and ionized gas, either in form of a nebula or a wind. To determine some physical parameters of the hydrogen line-forming regions, we used the code developed by \citet{Kraus2000} and modeled the spectra of the Pfund series of MN~83 and MN~112. For the computations we assume that the Pfund lines are optically thin and follow Menzel's case B recombination theory \citep{Menzel1938}. We fix the electron temperature at $T_{\rm e}=10\,000$~K \citep[a typical value for an ionized wind or nebula, see][]{Leitherer1991}. The shape of the emission lines can be represented with a Gaussian profile. The width of the lines of MN~112 is $\sim\,50\,$km\,s$^{-1}$ and hence of the order of the spectral resolution, but is smaller for the case of MN~83. The maximum observed member of the series (n$_{\rm max}$) is similar in both objects: Pf$_{38}$ for MN~83 and Pf$_{37}$ for MN~112. With those n$_{\rm max}$ values we derived hydrogen densities of $n_{\rm H}~=~(6.7\pm1.1)\times10^{13}$\,cm$^{-3}$ and $n_{\rm H}~=~(7.8\pm1.2)\times10^{13}$\,cm$^{-3}$ within the Pfund line-forming regions of MN~83's nebula and MN~112's wind, respectively.

\begin{deluxetable*}{lcccccc}
\tablecaption{Johnson-Cousins $BVR_{\rm C}I_{\rm C}$ photometry and IR colors (based on their 2MASS magnitudes) of our stellar sample. \label{table:photometry}}
\tablecolumns{5}
\tablewidth{0pt}
\tablehead{
\colhead{ID} &
\colhead{$B$} &
\colhead{$V$} &
\colhead{$R_{\rm C}$} &
\colhead{$I_{\rm C}$} &
\colhead{$J-H$} &
\colhead{$H-K_{\rm s}$} 
}
\startdata
MN~83  &...          		&...         		& $>$15.5$^a$    	&...				& 0.545$\pm$0.041 & 0.320$\pm$0.040\\
MN~108$^b$ & 18.430$\pm$0.343 	& 16.107$\pm$0.343 	& 14.823$\pm$0.048 	& 13.521$\pm$0.050 	& 2.241$\pm$0.062 & 1.132$\pm$0.045\\
MN~109 &...          		&...         		& $>$18.5$^c$    	&...				& 0.841$\pm$0.045 & 0.598$\pm$0.045\\
MN~112$^d$ & 17.054$\pm$0.099 	& 14.538$\pm$0.092 	& 12.710$\pm$0.029 	& 11.128$\pm$0.024 	& 3.096$\pm$0.067 & 1.695$\pm$0.057\\
\enddata
\tablenotetext{a}{fainter than $G_{\rm RP}=15.5$}
\tablenotetext{b}{$B=17.69$, $V=15.52$, $R_{\rm C}=15.53$ \citep{GKF2010}}
\tablenotetext{c}{fainter than $G_{\rm RP}=18.5$}
\tablenotetext{d}{$B=16.42$, $V=14.64$, $R_{\rm C}=13.68$ \citep{GKF2010}}
\end{deluxetable*}

The K-band spectrum of  MN~108 displays only a single but evident, apparently double-peaked emission line around $2.427~\mu$m. This line could correspond to a blend of two Si~{\sc iv} $\lambda\lambda2.4266,~2.4278~\mu$m lines, reported in emission in late-O supergiants by \citet{Lenorzer2002}, but because of the absence of other emission lines, we cannot identify it with confidence. Other than that, the IR-spectrum of MN~108 appears featureless.

MN~109 presents weak and wide absorption lines in the 2.41-2.47~$\mu$m range that correspond to the hydrogen Pfund series. To guide the eye, we over-plotted a synthetic spectrum of a star with $T_{\rm eff}=10\,000$~K and $\log g=3.0$, obtained using $synspec$ code \citep{Hubeny2011} with \citet{Kurucz1979} models. The spectrum shows weak and narrow Mg~{\sc ii} $\lambda\lambda2.4047,~2.4131~\mu$m lines in emission. The little emission features at $\sim2.31~\mu$m and $\sim2.37~\mu$m are residue from the telluric correction. 

Only two stars had optical counterparts so that we could obtain a spectrum at optical wavelengths. These are the stars MN~108 and MN~112. In Figure~\ref{fig:optical} we point out the position of the more intense identified lines, together with the position of the DIBs and telluric bands.

The optical spectrum of MN~108 shows lines of H~{\sc i}, He~{\sc i} and He~{\sc ii} in absorption, and resembles the one showed by \citet{GKF2010}. The presence of He~{\sc ii} lines indicates a spectral type O, in line with the results from the $K$-band spectrum. The absorption feature observed at $\lambda\sim5889$~\AA\ is most likely a blend of the He~{\sc i}~$\lambda5875.62$~\AA\ line with the Na~{\sc i} $\lambda\lambda$5889.95,~5895.92~\AA\ doublet.

The optical spectrum of MN~112 presents several strong emission lines, dominated by hydrogen and He~{\sc i}. It is practically the same as that obtained by \citet[][extremely similar to the spectrum of PCygni]{G2010}, and the equivalent widths of the main emission lines remain the same.

We would like to add, that none of the two stars displays emission of [O{\sc i}] in their optical spectra, which is one of the defining characteristics of B[e] stars.

\begin{figure*}
\centering
\includegraphics[width=0.521\columnwidth]{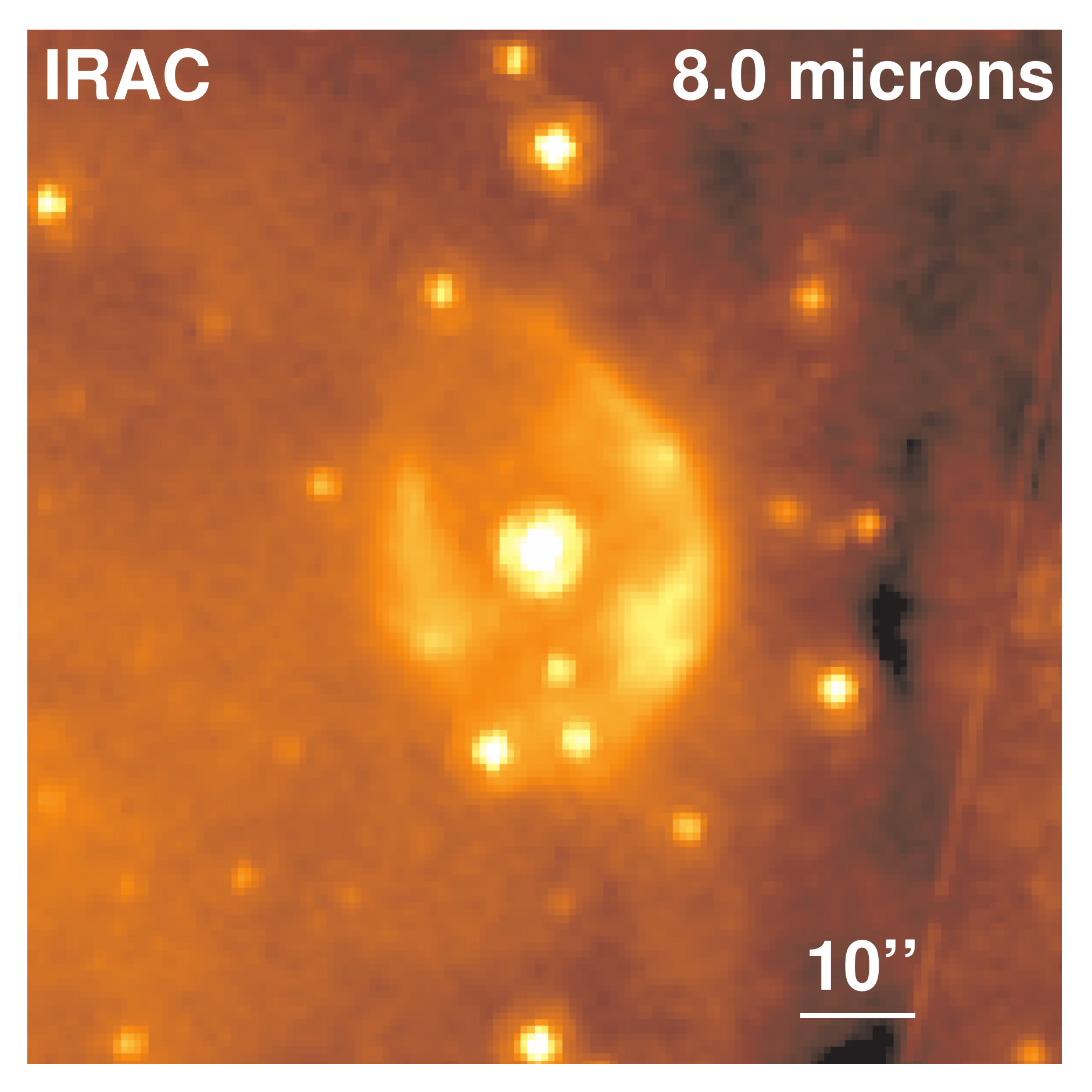}
\includegraphics[width=0.521\columnwidth]{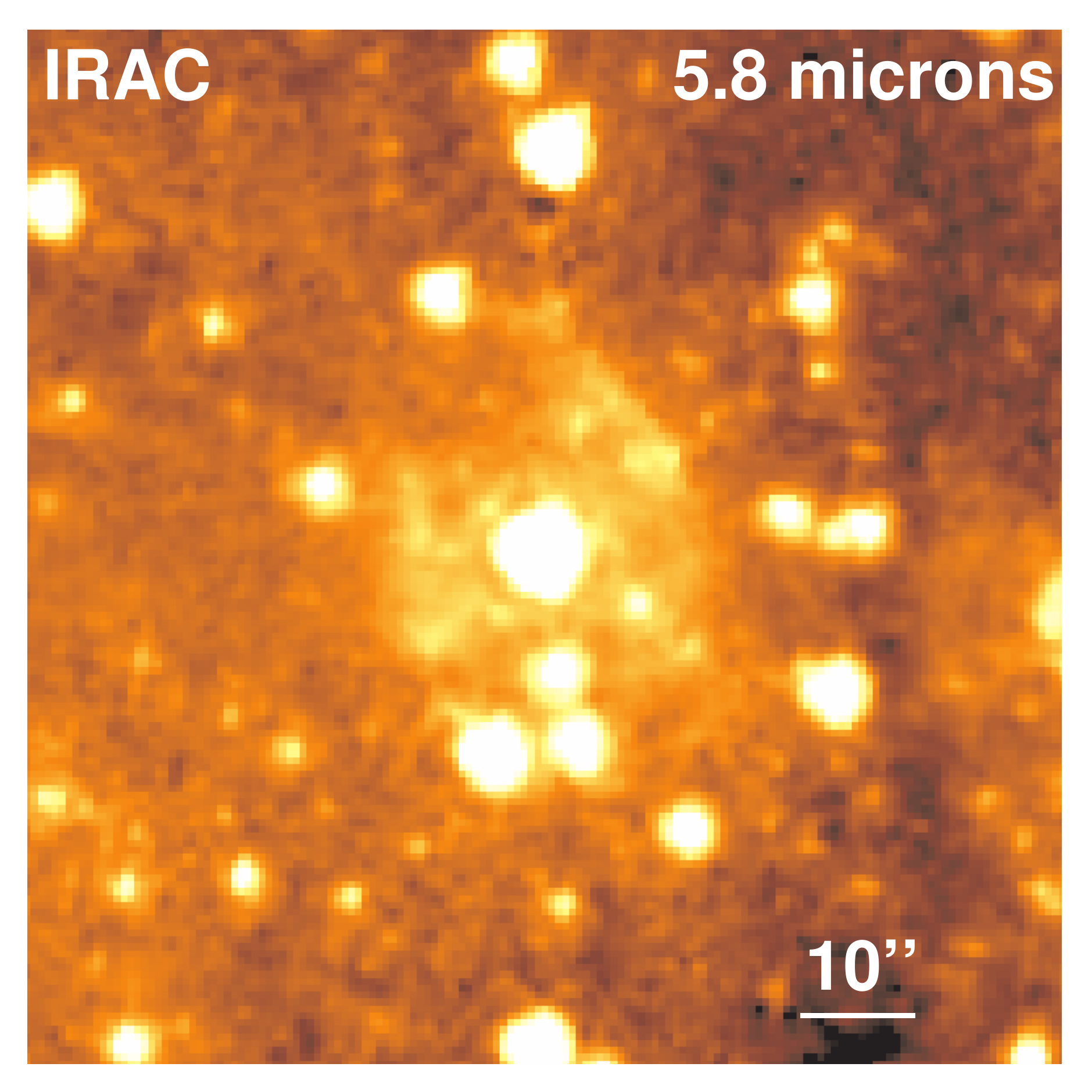}
\includegraphics[width=0.521\columnwidth]{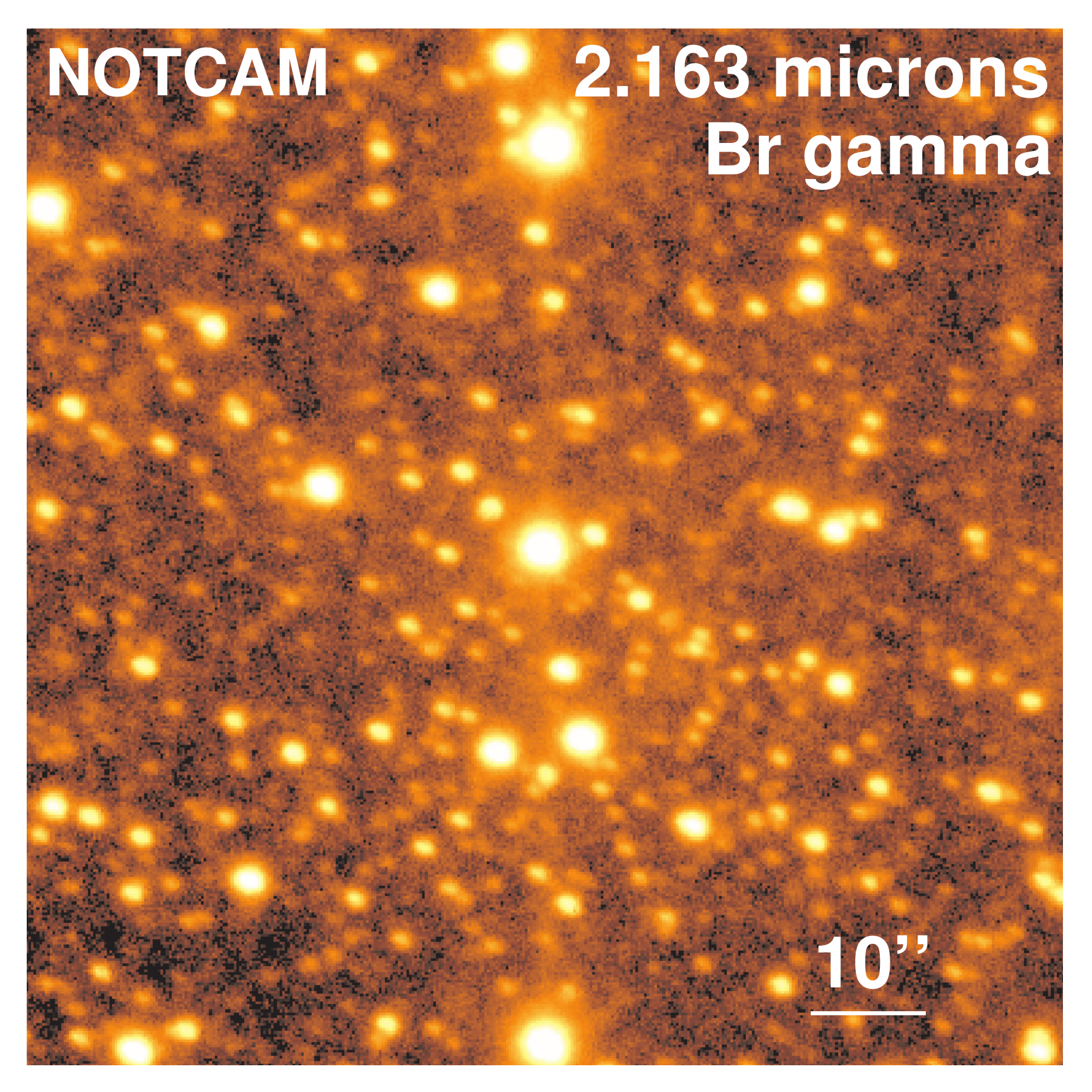}
\includegraphics[width=0.521\columnwidth]{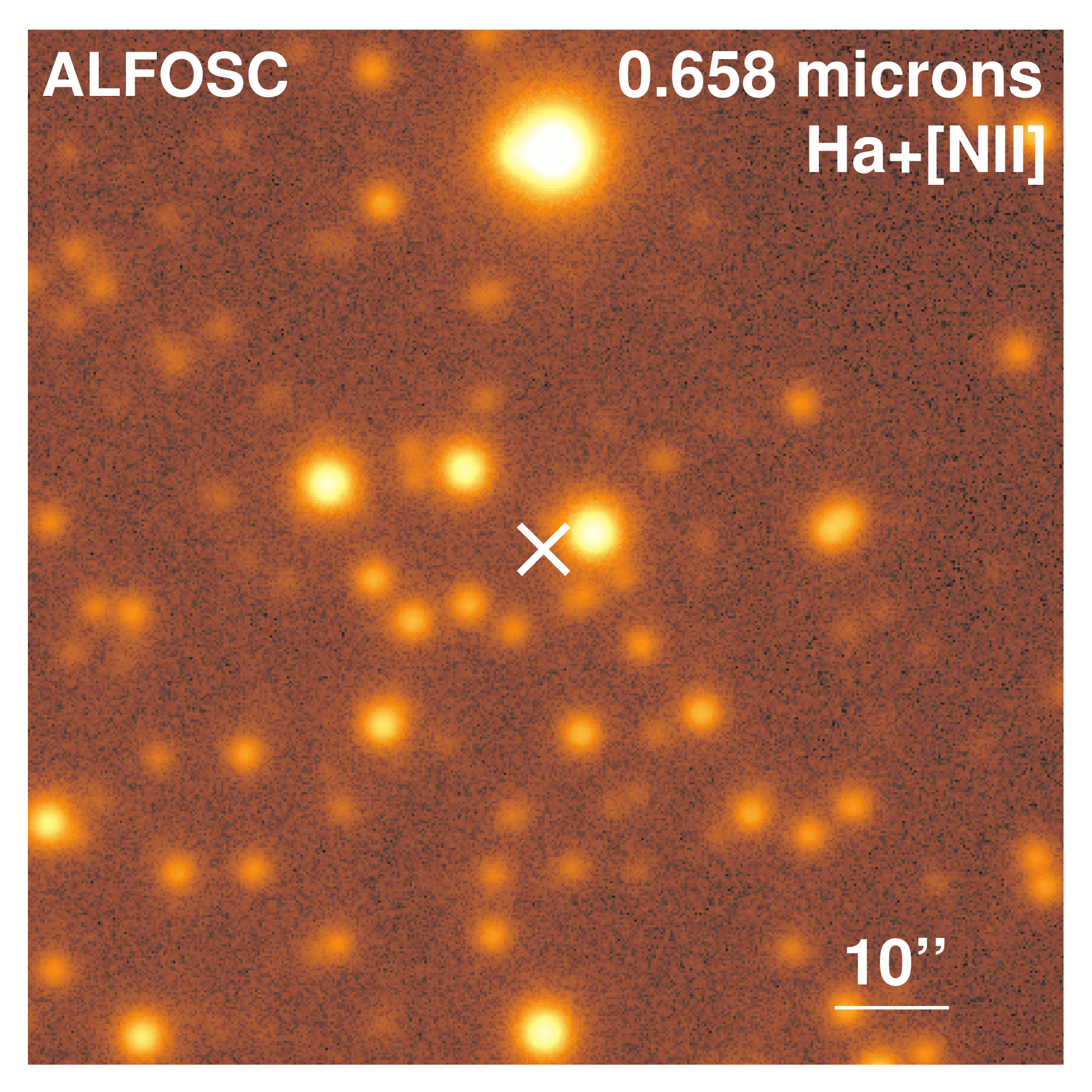}
\caption{Images of the MN~83 nebula. From left to right: Spitzer IRAC 8.0~$\mu$m, Spitzer IRAC 5.8~$\mu$m, NOTCAM Br$\gamma$, and ALFOSC H$\alpha$+[N\,{\sc ii}]. On the IR images MN~83 is the brightest star in the middle. In the optical range (ALFOSC) the position of MN~83 is marked with the white cross. The intensity levels on individual 
frames are arbitrarily chosen to maximize the visibility of the nebula. The FOV of all frames is $1\farcm5\times1\farcm5$. North is up, East is left.} 
\label{F-neb}
\end{figure*}

\begin{figure*}
\centering
\includegraphics[width=5.9cm]{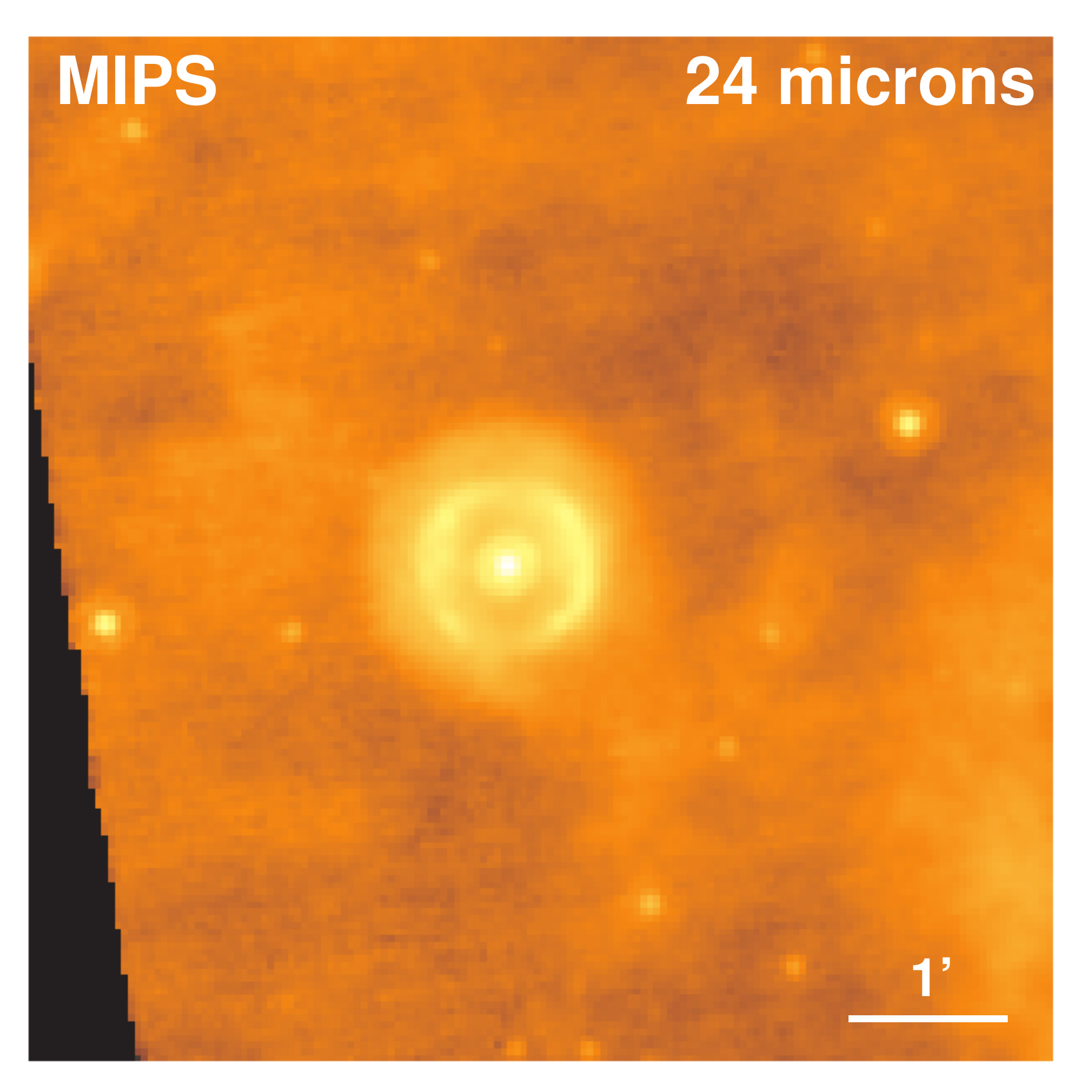}
\includegraphics[width=5.87cm]{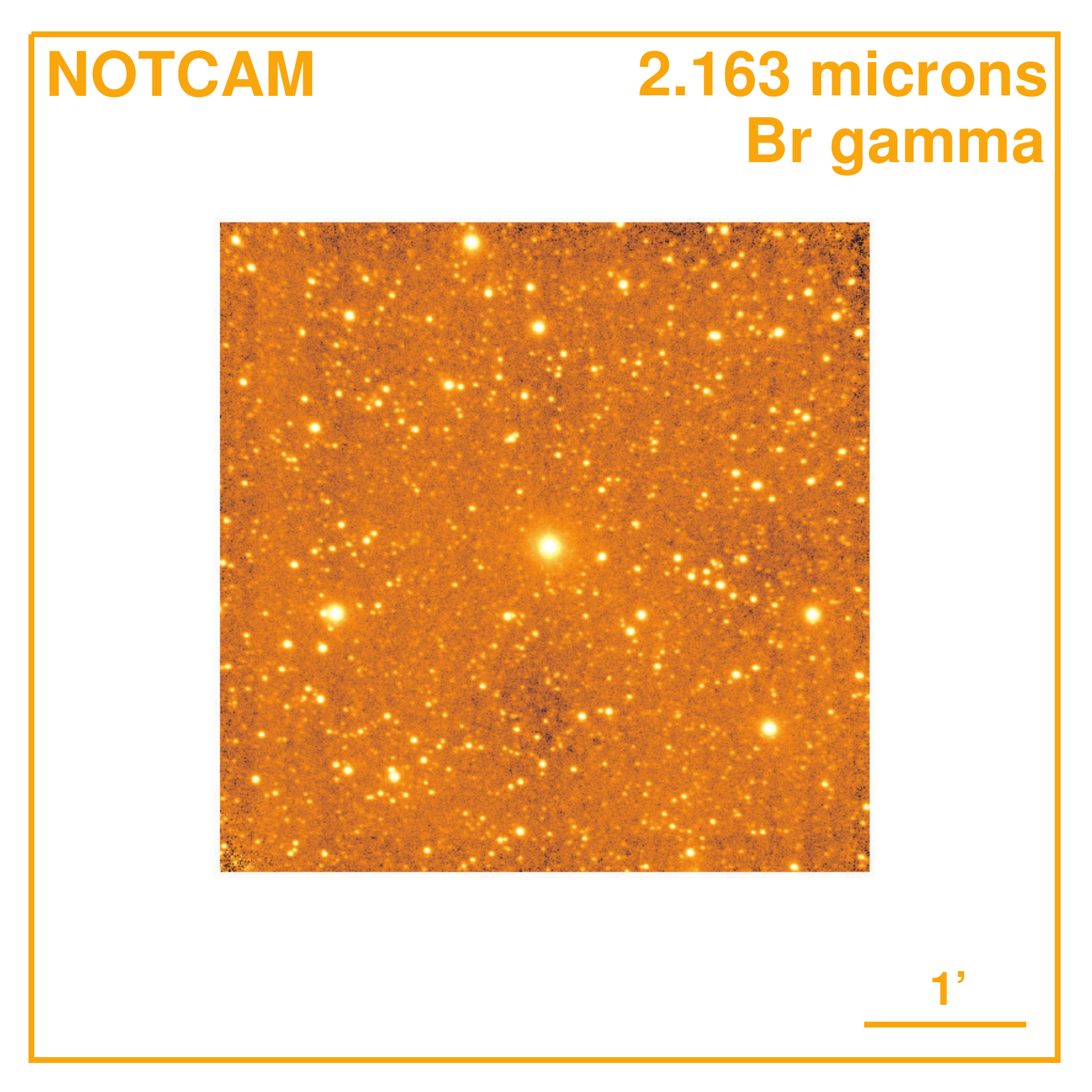}
\includegraphics[width=5.9cm]{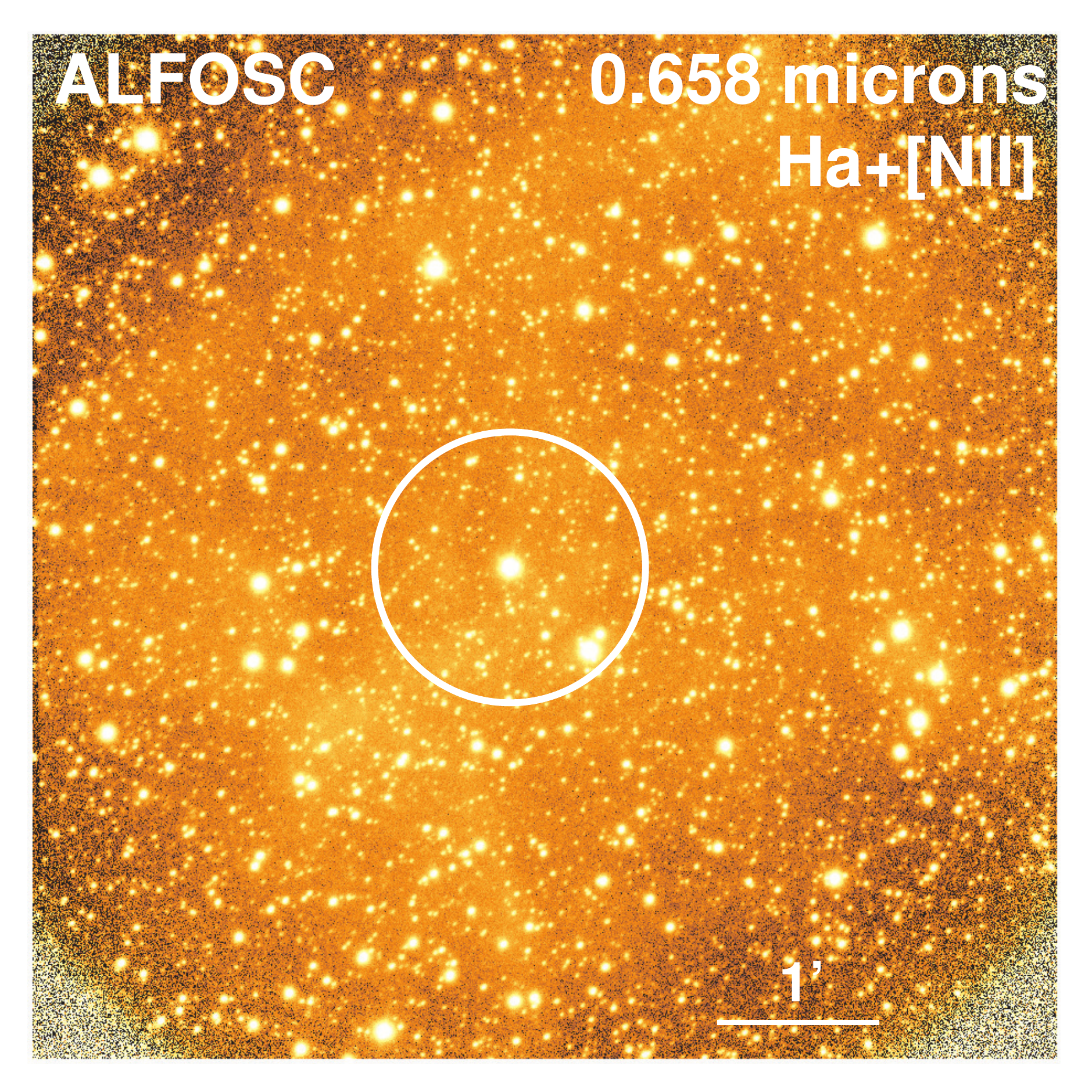}
\caption{Images of the MN~112 nebula. From left to right: Spitzer MIPS 24 $\mu$m, 
NOTCAM Br$\gamma$, and ALFOSC H$\alpha$+[N\,{\sc ii}]. 
The MN~112 is the brightest star in the middle. The intensity levels on individual 
frames are arbitrarily chosen to maximise the visibility of the nebulae. The white circle in the
ALFOSC image refers to the size and position of the circular nebula in 24 $\mu$m.
North up, East left. FOV of all frames is $6'.3\times6'.3$.} 
\label{F-neb2}
\end{figure*}

\subsection{Photometry}

Turning now to the photometric measurements, our observations confirm that MN~83 and MN~109 have no optical counterpart. Both stars remained undetected on our $BVR_{\rm C}I_{\rm C}$ images due to the possibly very high extinction toward these sources (see Sect.~\ref{Sect:color-color}). Through a comparison with the source Gaia DR2 4256466388864127104, which is the faintest source in the same field of view with a Gaia magnitude ($G_{\rm RP}$) measurement, we conclude that MN~83 is fainter than $G_{\rm RP} = 15.5$\,mag. In the field of view of MN~109, it is the source Gaia DR2 4322588299401926656 that allows us to conclude that our target is fainter than $G_{\rm RP} = 18.5$\,mag. These limits have been added to the measured magnitudes in Table~\ref{table:photometry}. 

For the other two objects, MN~108 and MN~112, we determined their magnitudes in the Johnson-Cousins $BVR_{\rm C}I_{\rm C}$ photometry. We wish to point out that our $I_{\rm C}$ magnitudes are the first measurements for both objects. The observed $B$ and $V$ magnitudes from MN~112 are in good agreement with those reported in APASS. We note, however, that our $BVR_{\rm C}$ magnitudes for both objects deviate by $0.5-1.0$\,mag from the NOMAD photometric values listed in \citet{GKF2010}. This could suggest that the objects are photometric variables. All values are included in Table~\ref{table:photometry}.

\subsection{Imaging}

The star MN~83 is particularly interesting, because its $K$-band spectrum indicates an ionized gas nebula in addition to the elliptic dusty nebula detected at 8~$\mu$m \citep{Davies2007} and at 24~$\mu$m \citep{GKF2010}. Therefore, we searched for further traces of the nebula at different wavelengths. First, we inspected all four bands of the IRAC frames at 3.6, 4.5, 5.8, and 8.0~$\mu$m. We found that the nebula is visible also at shorter wavelength than 8.0~$\mu$m, this being the frame where the nebula is brightest. In 5.8 and 4.5~$\mu$m the nebula gets progressively fainter until no detection at 3.6~$\mu$m. In Figure~\ref{F-neb} we present the 8.0 and 5.8~$\mu$m frames. The nebula has a semi-major axis in the North-South direction with clearly brighter filaments toward the East and West. It is evident that in the North, approximately at the position angle of 25$\degr$ (from North to East), the rim of the ellipse is broken or the material is significantly diluted so that the intensity of the emission dropped below the detection threshold compared to the rest of the nebula. The dimensions of the nebula at 8.0~$\mu$m are $44\arcsec\times28\arcsec$, slightly smaller than those at 24~$\mu$m. Without knowing the distance to MN83, no further calculations of e.g. its physical size or age can be done. 

Moving on to smaller wavelengths, we additionally opted to search for extended ionized gas structures in both Br$\gamma~2.16~\mu$m and H$\alpha$. The obtained images are included in Figure~\ref{F-neb}. No nebula has been detected at those wavelengths with the chosen total observing times of 30 min in H$\alpha$ and 7.5 min in Br$\gamma$.

The second object of high interest is MN~112 due to its previous classification as cLBV and its intense emission-line spectra both in the optical and near-infrared. Our optical image displays H$\alpha$+[N\,{\sc ii}] emission almost over the entire FOV (right panel in Fig.\,\ref{F-neb2}). 
However, in contrast to the sharp ring nebula seen at 24\,$\mu$m on the MIPS image (left panel in Fig.\,\ref{F-neb2}), the optical nebula appears to be smooth and with no clear substructure. Moreover, no pronounced emission was detected in the Br$\gamma$ line (middle panel in Fig.\,\ref{F-neb2}), nor was there any nebulosity or ring structure seen on the IRAC bands (not shown). 

Neither MN~109 nor MN~109 show any extended optical nebula.

\section{Discussion} \label{sec:dis}

As mentioned in Sect.~\ref{sec:intro} central stars of dusty ring nebulae might correspond to a variety of massive stars in transition phases, and in terms of blue supergiant central stars, B[e] supergiants and LBVs are the most likely candidates. In this section, we discuss the classification of our selected sample, based on their previous information and  results obtained from their optical and K-band spectra.

\subsection{CO molecular emission}
\label{Sect:CO}

None of the stars of our sample shows CO bands. Of all B[e]SGs with observed K-band spectra more than $65\%$ display CO band emission \citep{Kraus2019}. On the other hand, LBVs typically do not display CO band emission \citep[e.g.][]{Oksala2013}, with one exception, the LBV star HR~Car, which occasionally displays highly variable CO band emission \citep{Morris1997}. Therefore, the non-detection of warm CO emission in MN~83, MN~108, MN~109, and MN~112 is insufficient to distinguish between an LBV or a B[e]SG nature of the objects. Therefore, other characteristics should be considered.

\begin{figure*}[t!]
\begin{center}
\includegraphics[width=\textwidth]{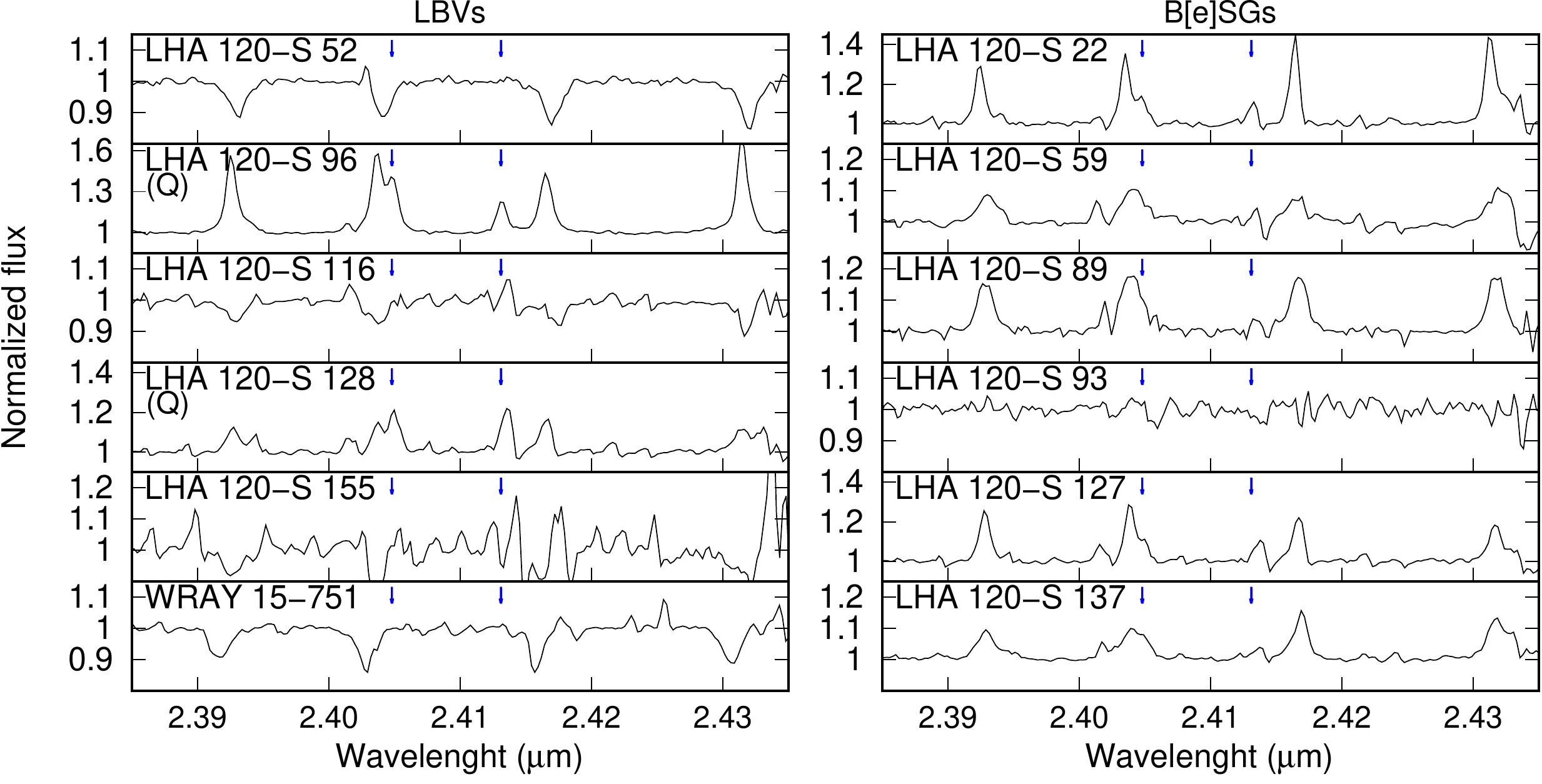}
\caption{Yet unpublished portion of the SINFONI spectra of the LBV and B[e]SG samples of \citet{Oksala2013}, highlighting the positions of the Mg~{\sc ii}  $\lambda\lambda2.4047,~2.4131~\mu$m lines. The ``Q'' label identifies the LBVs in quiescent state of their S\,Dor cycle.}
\label{fig:MgIIOksala}
\end{center}
\end{figure*}

\begin{figure*}[t!]
\begin{center}
\includegraphics[width=0.9\textwidth]{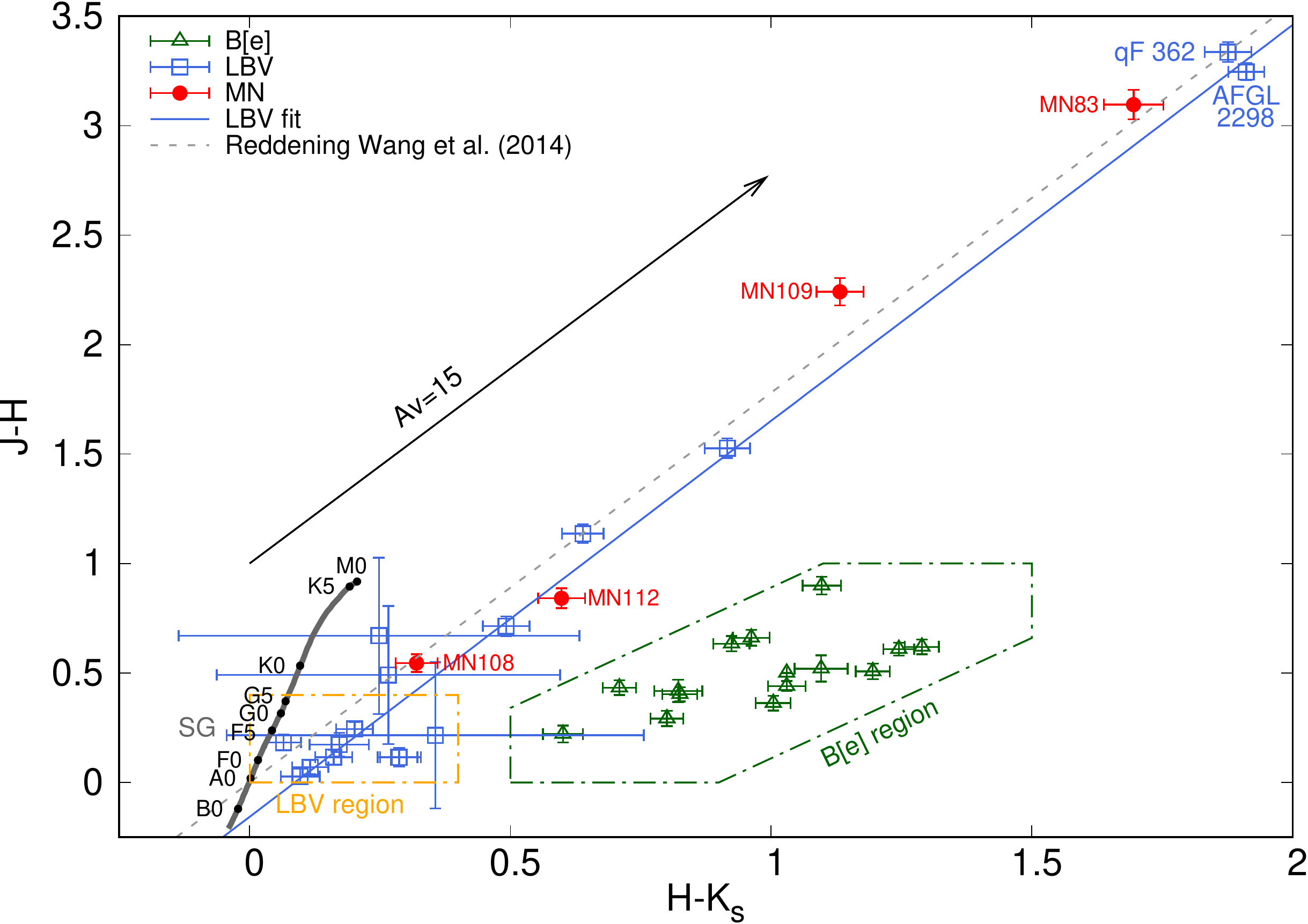}
\caption{$J-H$ versus $H-K_{\rm s}$ color-color diagram. LBV and B[e]SG regions from \citet{Oksala2013} are plotted in orange and green, respectively. LBVs are plotted with blue squares, and the blue continuous line is a linear fit to their positions. The reddening law obtained by \citet{Wang2014} is plotted as a dashed gray line, and a reddening vector for A$_{\text v}=15$ is included. The position of the normal supergiant (SG) stars from spectral-type O to M \citep{Worthey2011} are plotted in grey. The red circles mark the positions of our objects.\label{fig:JHK}}
\end{center}
\end{figure*}

\subsection{On the IR Mg~{\sc ii} emission lines}
\label{Sect:MgII}

As mentioned in Sect \ref{sec:lit}, MN~112 is considered a cLBV because its optical (low-resolution) spectrum resembles that of the confirmed LBV star P-Cyg, and because of the similarity of its nebula with GAL 079.29+11.46 \citep{G2010}. Hence, we could assume that MN~83 is also a cLBV because its IR spectrum displays many similarities to that of MN~112. Both objects present emission from the hydrogen Pfund series and from the Mg~{\sc ii} $\lambda\lambda2.4047,~2.4131~\mu$m lines, and in both stars the Mg~{\sc ii} lines are clearly more intense than the adjacent Pfund lines. In contrast, the star MN~109 shows weak and narrow  Mg~{\sc ii} lines in emission while the Pfund lines are in pure absorption, and MN~108 displays neither Pfund nor Mg~{\sc ii} lines.

To check if the lines of Mg~{\sc ii}  $\lambda\lambda2.4047,~2.4131~\mu$m are typically observed in emission in the spectra of LBV stars, we analyze the non-published spectral region around the Mg~{\sc ii} doublet in the SINFONI spectra of the sample of LBVs and B[e]SGs from \citet{Oksala2013} (see Figure~\ref{fig:MgIIOksala}). 

Of special interest for a comparison are hereby B[e]SGs with no CO band emission. Six B[e]SGs from the Magellanic Clouds were reported to display no CO band emission. Instead, all (except of LHA\,120-S\,93) show strong emission in the lines of the Pfund series \citep[see Figure~2 of][]{Oksala2013}. From the LBV sample two objects, LHA\,120-S\,96 (= S\,Dor) and LHA\,120-S\,128, are in quiescent state of their S\,Dor cycle and, therefore, display a very similar emission spectrum as the B[e]SGs. The other four LBVs appear to be in their outburst state, in which they display mostly absorption-line spectra resembling cool(er) supergiants. In these stars, the Pfund lines appear in absorption \citep[see Figure~3 of][]{Oksala2013}.

We inspected the unpublished spectral ranges (longwards $2.4~\mu$m) of these two samples of objects. They are displayed in Figure~\ref{fig:MgIIOksala}. The left panel contains the LBVs, and the right panel the B[e]SGs. The positions of the Mg~{\sc ii} $\lambda\lambda2.4047,~2.4131~\mu$m lines are marked by the arrows. From this figure, we find that all B[e]SGs, except LHA\,120-S\,93, have the Mg~{\sc ii} lines in emission, and in all of them, the intensity in the Mg~{\sc ii} lines is considerably lower than in the adjacent Pfund lines. For the LBVs, we can see that the two objects in quiescence also have the Mg~{\sc ii} lines in emission, and in one of them (LHA\,120-S\,128), the Mg~{\sc ii} 
lines are more intense than the Pfund lines, similar to our two cLBV objects MN~83 and MN~112. Interestingly, at least three of the LBVs in outburst (LHA\,120-S\,116, LHA\,120-S\,155, WRAY 15-751) show indication for emission in the Mg~{\sc ii} lines, similar to our object MN~109. This suggests, that MN~109 might be an LBV in a current outburst state. The rather low effective temperature, pointing toward an A-type supergiant star, seems to support such a classification.

LBVs have already been reported to display Mg~{\sc ii} $\lambda\lambda2.4047,$ $2.4131~\mu$m lines in emission. In their infrared atlas of early-type stars, \citet{Lenorzer2002} present the spectra of three  LBVs ($\eta$~Carinae, AG~Carinae and P~Cygni) showing these features (see their Figure 6). Moreover, the spectrum of P Cygni shows Mg~{\sc ii} emission lines stronger than Pfund emission lines.

\citet{Clark2018} also presented K-band spectra for early-type stars of the Quintuplet cluster. They claimed that the Mg~{\sc ii} lines in emission become prominent for spectral type B3 and later. Particularly, the three LBVs found in the cluster show Mg~{\sc ii} $\lambda\lambda2.1369,~2.1432~\mu$m lines in emission. Only one of these spectra covers the region longwards $2.4~\mu$m and it also shows emission in  Mg~{\sc ii} $\lambda\lambda2.4047$, $2.4131~\mu$m lines (see their Figures 1 and 7).

Of course, our rather qualitative analysis, based on just a small sample of stars, requires to be repeated in more depth 
and with a statistically significant sample of stars. Nevertheless, the intensity of the IR Mg~{\sc ii} lines seems 
to be related to the density/extent of the circumstellar material and could be considered as a complementary criterion to classify stars as LBVs.

\subsection{$J-H$ versus $H-K_{\rm s}$ color-color diagram}
\label{Sect:color-color}

With the exception of the subtle distinction about the intensity of the IR emission lines of Mg~{\sc ii} between LBV and B[e]SG stars highlighted in the previous Section, the $K$-band spectra of quiescent LBVs are practically indistinguishable from those of B[e]SGs with no CO bands. However, these two groups of stars fall in different regions of a $J-H$ vs $H-K_{\rm s}$ color-color diagram \citep{Oksala2013}, since the near-IR excess observed in B[e]SG arises from the warm/hot circumstellar dust, while the one in LBVs comes from free-free emission due to stellar winds. Thus, the IR color-color diagram can be used as additional diagnostics to discriminate between B[e]SGs and LBVs.

In Figure~\ref{fig:JHK} we plot the IR color-color diagram and indicate the corresponding regions for LBVs (orange box) and B[e]SGs (green box) taken from \citet{Oksala2013}. There, we plot with green triangles and blue squares the B[e] sample from \citet{Oksala2013} and the LBV stars listed in \citet{Clark2005}, \citet{AAdland2019} and \citet{Campagnolo2018}. The blue solid line is a linear fit of the position of confirmed LBVs in the $J-H$ vs $H-K_{\rm s}$ diagram. This linear fit is in excellent agreement with the reddening law obtained by \citet[][plotted in a dashed gray line]{Wang2014}. This law has been constructed from a sample of giant stars based on the NIR spectroscopic survey project APOGEE, by finding the relations between three NIR intrinsic colors and the effective temperature by fitting the bluest colors with a quadratic line. The red circles mark the positions of our four objects in this color-color diagram based on their 2MASS magnitudes (included in Table~\ref{table:photometry}). All of them are distributed along the traced reddening curve and in line with confirmed LBVs, implying that, depending on their intrinsic reddening values, their unreddened colors would place them into the box of confirmed LBVs or on the SG sequence.

Using the $E(B-V)$ values from the literature and assuming a total-to-selective extinction ratio $R_{\text V}=A_{\rm V}/E(B-V)=3.1$ \citep{Schultz1975}, we obtained the corresponding extinction values $A_{\rm V}$ for the LBVs. For the two LBVs in the upper-right corner of the diagram, we obtained $A_{\rm V}=24.8$ (for the upper, qF~362) and $A_{\rm V}=27.9$ (for the lower, AFGL~2298). With these values, we estimated the extinctions for our four objects. If we consider that MN~83 and MN~112 are cLBVs, their de-reddened $J-H$ versus $H-K_{\rm s}$ values ​​should fall in the LBV region. With that restriction, we found lower limits of $A_{\rm V}~\sim~23$ for MN~83 and $A_{\rm V}~\sim~4$ for MN~112. In a similar way, if we de-redden the position of MN~109 and MN~108 in the color-color diagram with the lower limit of $A_{\rm V}$, i.e., when they cross the threshold of the LBV domain, we obtain $A_{\rm V}~\sim~16$ and $A_{\rm V}~\sim~5$, respectively. In the case of MN~109 the needed extinction is lower than that given by \citet{PhillipsRamos2008}. The high extinction values obtained for MN~83 and MN~109 are consistent with the fact that both objects are not observed in the optical range. 

\section{Conclusions}
\label{sec:con}

We analyze four objects, which have dusty envelopes resolved on {\it Spitzer} 24~$\mu$m images: [GKF2010] MN~83, [GKF2010] MN~108, [GKF2010] MN~109 and [GKF2010] MN~112. These objects have been suggested to host either a cLBV or a BSG central star. 

To improve our knowledge on such massive stars in transition phases and on their surroundings, we obtained the first medium resolution K-band spectra in the $2.3\,-\,2.47\,\mu$m region for these four objects, because this spectral region provides suitable information that can be used to classify evolved massive stars. For MN~108 and MN~112 we supplement the IR spectra with optical spectra and photometric $BVR_{\rm C}I_{\rm C}$ measurements, as only these two stars have optical counterparts. We report the first $I_{\rm C}$-band magnitudes of $13.521\pm0.050$\,mag for MN~108 and $11.128\pm0.024$\,mag for MN~112. We derived lower limits of $A_{\rm V} > 16$\,mag and $A_{\rm V} > 23$\,mag for MN~109 and MN~83, respectively, confirming their non-detections at optical wavelengths.

From our optical imaging, we found that none of the stars shows H$\alpha$ emission comparable to the infrared ring nebulae reported in the literature. However, MN~112 seems to be surrounded by extended,
diffuse H$\alpha$ emission, with approximately 2.5 times the size of the infrared ring nebula 
(Fig.~\ref{F-neb2}). Unfortunately, our Br$\gamma$ images of MN~83 and MN~112 also do not show any extended emission, whereas clear emission from their dusty nebula is seen at 8~$\mu$m and 5.8~$\mu$m (Fig.~\ref{F-neb} for MN~83) and at 24~$\mu$m (Fig.~\ref{F-neb2} for MN~112).

MN~83 and MN~112 show characteristics typically seen in LBV stars. The $K$-band spectra of both objects display intense IR Mg~{\sc ii} lines in emission along with emission of the Pfund series and they lack CO band emission. Their Mg~{\sc ii} lines are more intense than the adjacent Pfund lines. Such a behavior is also seen in other LBV stars (see Sect. \ref{Sect:MgII}). The optical spectrum of MN~112 displays no [O~{\sc i}] emission, which would otherwise point to a B[e]SG classification. Moreover, the unreddened JHK colors place both stars in the loci of LBVs, so that a cLBV classification of both stars seems justified. One might speculate whether the dusty ring nebulae of these two objects might be remnants of either a giant eruption or S~Dor cycle(s).

The situation for the other two objects is less clear. MN~109 seems to be an A-type supergiant star with $T_{\rm eff} \le 10\,000$~K and $\log\,g \simeq 3$. Its $K$-band spectrum displays the Pfund lines in absorption along with weak emission of the IR Mg~{\sc ii} lines, similar to LBVs in outburst. Hence, MN~109 could be considered as a cLBV star in its active phase.

The optical spectrum of MN~108 reveals narrow photospheric lines of H~{\sc i}, He~{\sc i} and He~{\sc ii} characteristic for a hot, O-type supergiant. The K-band spectrum displays only Si~{\sc iv} emission, in agreement with the classification of an O-type supergiant, but is otherwise featureless. The absence of pronounced wind lines in the optical spectrum might suggest that the star is a regular, non-LBV object. Whether the dusty envelope (the witness of a phase of intense mass loss),  might still be a remnant from a prior red-supergiant phase can be neither confirmed nor excluded, but requires further investigation.

\acknowledgments

We thank the anonymous referee for the valuable comments that helped us to improve this work.

This research made use of the NASA Astrophysics Data System (ADS) and of the SIMBAD database, 
operated at CDS, Strasbourg, France.

This research was made possible through the use of the AAVSO Photometric All-Sky Survey (APASS), funded by the Robert Martin Ayers Sciences Fund and NSF AST-1412587, and of the NASA/IPAC Infrared Science Archive, which is funded by the National Aeronautics and Space Administration and operated by the California Institute of Technology.

The ALFOSC instrument is provided by the Instituto de Astrofisica de Andalucia (IAA) under a joint  agreement with the University of Copenhagen and NOTSA. 

We thank Shane Moran and Joonas Viuho for obtaining the NOTCam images.

M.K. acknowledges financial support from the Grant Agency of the Czech Republic (GA \v{C}R, grant number 20-00150S). The Astronomical Institute Ond\v{r}ejov is supported by the project RVO:67985815. M.L.A. thanks financial support from the University of La Plata (11/G160) and L.S.C. acknowledges financial support from CONICET (PIP 0177) and Agencia (PICT 2016-1971). T.E. acknowledges financial support from Estonian Science Foundation institutional research funding IUT40-1 of the Estonian Ministry of Education and Research.

This project has received funding from the European Union's
Framework Programme for Research and Innovation Horizon 2020 (2014-2020) under the Marie Sk\l{}odowska-Curie Grant Agreement No. 823734, and from the European Union's Horizon 2020 research and innovation programme under grant agreement No 730890. This material reflects only the authors views and the Commission is not liable for any use that may be made of the information contained therein.

%

\vspace{5mm}
\facilities{Gemini:Gillett(GNIRS), NOT, VLT:Yepun}


\software{IRAF \citep{IRAF1986}
          }



\bibliography{biblio}{}
\bibliographystyle{aasjournal}



\end{document}